\documentclass{IEEEtran}

\usepackage[table]{xcolor}
\usepackage{cite}
\usepackage{graphicx}
\usepackage{stfloats}
\usepackage{amssymb}
\usepackage{pifont}
\usepackage{booktabs}
\usepackage{threeparttable}
\usepackage{footmisc}
\usepackage{array}
\usepackage{multirow}
\usepackage{enumitem}
\usepackage{url}
\usepackage{tabularx}
\graphicspath{{Figures/}}
\DeclareGraphicsExtensions{.pdf, .png, .eps}

\begin{document}

\title{Control-Compute Governance in Agentic AI-RAN: AI Agents as Both Controllers and Workloads}

\author{
	Le Xia,~\IEEEmembership{Member,~IEEE},
    Rose Qingyang Hu,~\IEEEmembership{Fellow,~IEEE},
    Yuan Zhou,~\IEEEmembership{Member,~IEEE},\\
    and Haijian Sun,~\IEEEmembership{Senior Member,~IEEE}
\thanks{

Le Xia, Rose Qingyang Hu, and Yuan Zhou are with Virginia Tech, USA;

Haijian Sun is with University of Georgia, USA.

}}
\maketitle
\bstctlcite{BSTcontrol}

\begin{abstract}
Recent advances in artificial intelligence-radio access network (AI-RAN) are placing large language model (LLM)-driven agents within the open RAN (O-RAN) control hierarchy.
A promising deployment for this agentic AI-RAN co-locates LLM inference with physical-layer (PHY) communication processing over the same accelerated compute pool for infrastructure sharing, data locality, and low control latency.
However, this co-location induces bidirectional control-compute coupling, as the agent competes with PHY processing for compute, while its reasoned gNB control actions may alter future PHY workload and hence the compute available to its next inference.
To this end, this article proposes the control-compute contract, a coordination protocol linking AI-RAN workload governance with O-RAN control execution.
Specifically, we first expose direct compute contention on an over-the-air (OTA) O-RAN testbed, where continuous LLM inference increases the PHY decoding time approximately eightfold.
We then formulate the contract as three protocol rules with operator-requirement protection, and map it onto O-RAN and AI-RAN functions as a five-stage workflow.
Afterward, an OTA-calibrated case study of simulated uplink gNB control indicates that the contract keeps the proportion of slots violating the PHY time budget below $0.9\%$, against $20.7\%$ under ungoverned co-location, while achieving faster gNB control than a fixed compute reservation benchmark.
Finally, we identify several open issues and outlooks for its practical deployment.

\end{abstract}

\section{Introduction}
Artificial intelligence-radio access network (AI-RAN) is garnering increasing attention as a promising wireless architecture for converging AI with the RAN.
The AI-RAN Alliance defines its three pillars, namely AI-for-RAN to improve RAN functions with AI, AI-and-RAN to host AI and RAN workloads on common infrastructure, and AI-on-RAN to deliver AI services over the RAN~\cite{airanarch}.
As these pillars converge toward 6G, the gNB is no longer a fixed-function appliance but an accelerated compute node whose capacity is shared among physical-layer (PHY) processing, RAN intelligence, and AI applications.
Within this vision, large language model (LLM)-driven AI agents are being explored to automate intelligent gNB control, giving rise to a paradigm of~\textit{agentic AI-RAN}~\cite{sun2026sc3}.

Nevertheless, implementing agentic AI-RAN within conventional deployments may be strictly constrained by proprietary interfaces from single vendors to acquire RAN observation data and issue gNB control actions~\cite{polese2023understanding}.
Fortunately, open RAN (O-RAN) offers flexible interfaces and open platforms for different vendors' applications, forming a natural architectural basis for AI agents, whose control logic can perfectly reside in xApps on the near-real-time RAN intelligent controller (Near-RT RIC) for programmable gNB control~\cite{oranoad}.
Their LLM inference can further run alongside the gNB's accelerated PHY processing over a shared compute pool to share infrastructure, preserve data locality, and reduce control latency.

In parallel, this compute co-location approach introduces a bidirectional control-compute coupling issue for agentic AI-RAN.
Concretely, the LLM inference workload competes with PHY processing for compute, while the agent's gNB control actions may alter the PHY workload and further affect the compute available to its subsequent inference.
Note that such coupling goes beyond allocating compute between the two concurrent workloads, as an inference allocation compatible with the current PHY workload may become incompatible once the agent's control action executes.
Conversely, deferring or interrupting inference to protect PHY processing can leave the agent without a new control action.
Hence, compute allocation and control execution cannot be handled separately, and governing this coupling presents three interrelated challenges.

\textbf{Dynamic LLM Compute Adjustment for PHY Protection:}
Allocating compute to LLM inference requires estimating its impact on future PHY processing time under the current system workload.
Further, given that physical RAN states and operator requirements may change at any time, the inference compute must be adjusted dynamically to meet PHY time budgets without compromising the agentic control quality.

\textbf{Sustaining gNB Control under LLM Inference Interruption:}
Governing the compute contention may require deferring or even interrupting LLM inference for PHY protection, preventing the agent from issuing timely gNB control actions.
However, simply reusing an earlier gNB control action is also insufficient, since its validity depends on the earlier RAN state and operator requirements that may have changed.
Thus, valid gNB control must be sustained without fresh LLM outputs.

\textbf{Compute Compatibility Across gNB Control Execution:}
An issued gNB control action may not take effect immediately, and compute allocation should remain compatible with current PHY processing, without assuming that the new PHY workload is already active or that it continues to rely on the previous workload after application.
Consequently, the compute adjustment and action execution must be properly sequenced during the gNB configuration transition.

Existing AI-RAN research has primarily explored either AI agent-driven gNB control~\cite{elkael2025agentran, chatzistefanidis2025symbiotic} or compute allocation between AI and RAN workloads~\cite{yet2026slallm, schiavo2024yyr, schiavo2024cloudric}.
However, to the best of our knowledge, none of them explicitly coordinate the two together with bidirectional coupling, and current AI-RAN and O-RAN specifications also leave this issue open~\cite{airanarch, oranaal}.
To this end, this article proposes the control-compute contract, a coordination protocol that links gNB control verification with LLM inference admission, requires compute compatibility before control action execution, and determines subsequent inference admission under the projected PHY workload.
In short, our main contributions are summarized as follows.
\begin{itemize}
\item We first characterize two control-compute coupling paths in agentic AI-RAN, through which separately verified gNB control actions and LLM inference can become incompatible.
	Empirically, our over-the-air (OTA) experiments on an O-RAN testbed substantiate the compute contention in the first coupling path, showing that continuous LLM reasoning on one shared graphics processing unit (GPU) increases the PHY decoding time eightfold.
\item We then design the control-compute contract around three protocol rules and an operator-requirement protection condition, and map it onto O-RAN and AI-RAN functions as a five-stage workflow, which coordinates gNB control verification and E2 execution at the Near-RT RIC with LLM-PHY compute allocation at the gNB host.
\item Afterward, we evaluate the contract in an OTA-calibrated case study of simulated uplink gNB control.
	The results show a PHY time-budget overrun proportion of below $0.9\%$ against $20.7\%$ using the ungoverned co-location benchmark, while realizing up to four times faster gNB control than the worst-case reservation benchmark.
	We further compare our contract with two execution-order alternatives during gNB configuration changes to illustrate the benefit of coordinating inference adjustment and admission with gNB control execution.
	Finally, several open issues with the prospects of this work are outlined.
\end{itemize}
 
\begin{figure*}[t]
\centering
\includegraphics[width=0.995\linewidth]{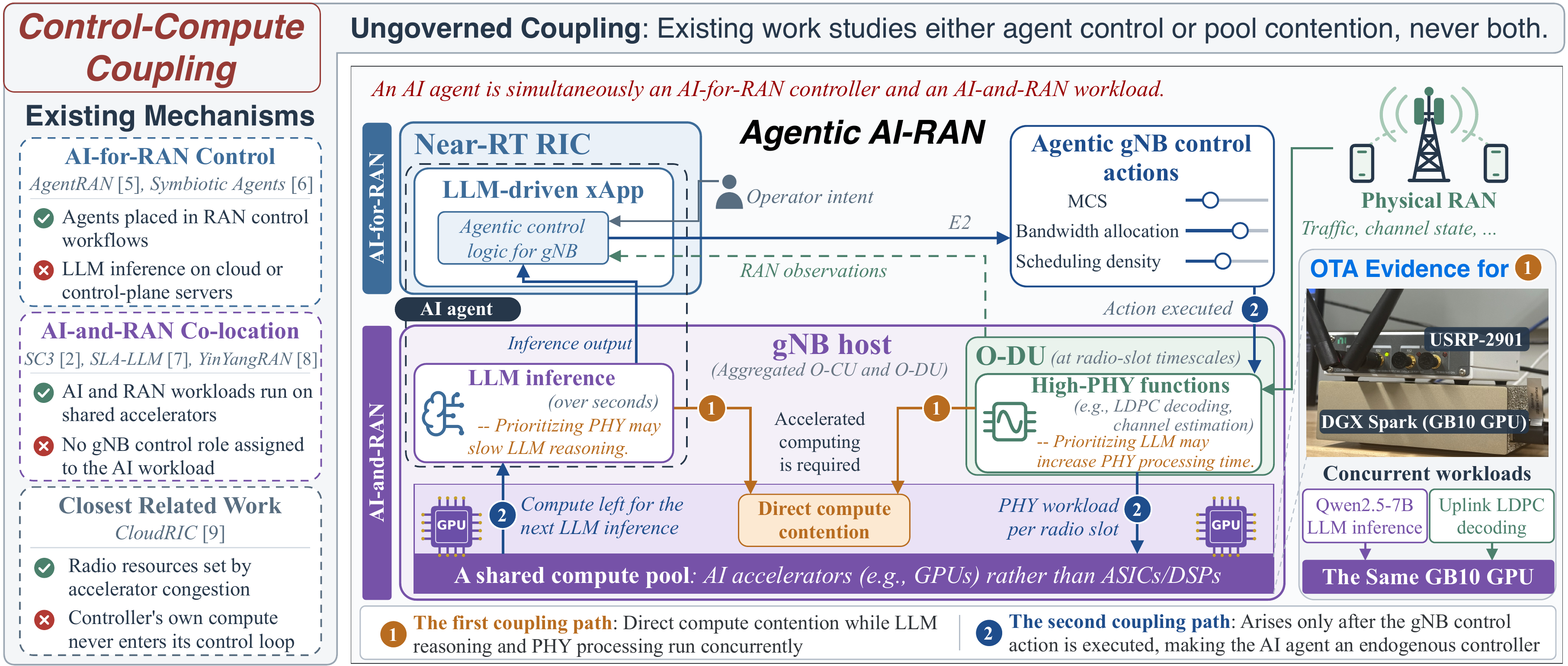}
\caption{Agentic AI-RAN with two control-compute coupling paths (center), alongside existing mechanisms (left) and the OTA testbed for the first path (right).}
\label{fig:loops}
\end{figure*}
\section{Ungoverned Control-Compute Coupling in Agentic AI-RAN}
\subsection{Two Coupling Paths over a Shared Compute Pool}\label{sec:coupling path}
Agentic AI-RAN sits at the intersection of the AI-for-RAN and AI-and-RAN pillars~\cite{kundu2026airan, airanarch}, where AI agents are simultaneously AI-for-RAN controllers and AI-and-RAN workloads.
O-RAN offers a concrete architectural realization, where the agents are logically situated in the rApp and xApp roles of its RIC hierarchy as part of the gNB control path~\cite{elkael2025agentran}.
Meanwhile, O-RAN specifications permit an implementation to aggregate the Near-RT RIC, O-RAN central unit (O-CU), and O-RAN distributed unit (O-DU)~\cite{oranoad}, allowing the agent's LLM inference to run on the gNB host.

The agent's dual role, however, opens two control-compute coupling paths, as depicted in Fig.~\ref{fig:loops}.
The first arises from direct contention in the shared pool.
On the gNB side, the O-DU executes latency-sensitive high-PHY functions, such as uplink low-density parity-check (LDPC) decoding and channel estimation, at radio-slot timescales.
In AI-RAN, such functions are designed to run on AI accelerators (e.g., GPUs)~\cite{kundu2026airan}, rather than on traditional application-specific integrated circuits (ASICs)/digital signal processors (DSPs).
In parallel, the AI agent on the RIC side operates on shared accelerators and turns RAN observations and operator intent into a candidate gNB control action over seconds~\cite{chatzistefanidis2025symbiotic}.
Such workload overlap may extend PHY processing time, while prioritizing PHY execution slows the agent's reasoning~\cite{schiavo2024yyr}.
The contention intensity also varies with traffic, channel state, and RAN configuration, which shape the PHY workload presented to the pool.
Furthermore, since it can also occur during LLM inference, the first coupling path persists whether or not the resulting candidate gNB control action is executed.

By contrast, the second coupling path arises only after the control action is executed.
For example, an action that changes the modulation and coding scheme (MCS), the bandwidth allocation, or the scheduling density can set the PHY workload per radio slot~\cite{schiavo2024cloudric}, subsequently altering the compute left for the next-round LLM inference.
Hence, this closes a feedback path from the agent, through the gNB's accelerated PHY processing, and back to the agent's subsequent reasoning.
Moreover, such feedback makes each agent an endogenous controller whose available compute is set inside the control loop it runs rather than being provisioned from outside.

\subsection{Gaps in Existing Control and Compute Mechanisms}
As summarized on the left of Fig.~\ref{fig:loops}, existing AI-RAN research falls into two classes, neither of which considers an agent that controls the gNB while reasoning on its pool.
One class studies AI-for-RAN control yet keeps LLM reasoning outside the gNB's compute pool.
For instance, AgentRAN~\cite{elkael2025agentran} and Symbiotic Agents~\cite{chatzistefanidis2025symbiotic} ran the LLM inference on cloud services or dedicated control-plane servers, adding round-trip latency to every control decision.
The other studies AI-and-RAN co-location without examining this bidirectional control-compute coupling.
For example, SC3~\cite{sun2026sc3} and SLA-LLM~\cite{yet2026slallm} co-locate LLM inference with live 5G processing under fixed GPU partitions, while YinYangRAN~\cite{schiavo2024yyr} multiplexes PHY and generic machine-learning workloads.
In fact, CloudRIC~\cite{schiavo2024cloudric} comes closest to bridging the two sides, as it allocated radio resources in response to accelerator congestion.
Even there, its controller runs as a lightweight CPU thread outside the shared pool, and its own decision-making compute never enters its control loop.
The same separation remains in current AI-RAN architecture documents~\cite{airanarch} and O-RAN specifications~\cite{oranaal}.

\subsection{OTA Evidence of Control-Compute Coupling}\label{sec:ota}
The direct LLM-PHY contention described above is first put to an OTA test, which targets only the first coupling path, since the action-to-workload link of the second is already evidenced in~\cite{schiavo2024cloudric}.
Specifically, the OTA deployment on the right of Fig.~\ref{fig:loops} comprises an NVIDIA DGX Spark, an OpenAirInterface gNB with aggregated O-CU and O-DU functions running the uplink LDPC decoding on a GB10 GPU, and a Near-RT RIC connected over E2.
Meanwhile, a Qwen2.5-7B LLM~\cite{qwen2025}, resident on the same GPU and served by llama.cpp, runs as the co-located AI workload to reason continuously over gNB control tasks.
In addition, a commercial 5G module is wirelessly attached through a USRP-2901 front end with a configured $1$-Mbps uplink rate, and the GPU time consumed in decoding an uplink LDPC code block serves as the PHY processing-time metric.
The test spans $19$ consecutive $90$-s time frames, and each places the GPU under a certain test condition, covering \textit{no LLM loaded}, \textit{LLM loaded but not reasoning}, an \textit{emulated GPU load of $25\%$} or \textit{$75\%$} in place of the LLM, and \textit{LLM reasoning continuously}.
Baseline frames with the LLM loaded but not reasoning bracket each frame with added GPU load to check that the decoding time recovers.

\begin{figure}[t]
\centering
\includegraphics[width=\linewidth]{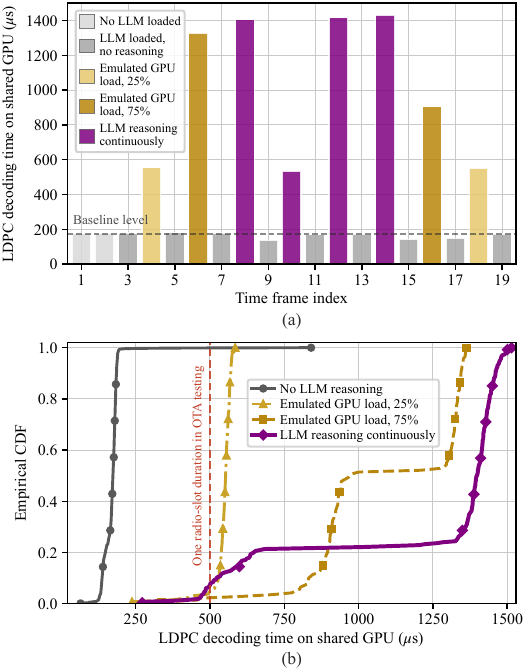}
\caption{OTA evidence of LLM-PHY compute coupling on one GPU under five test conditions. (a) Per-frame LDPC decoding times and (b) their CDFs.}
\label{fig:trace}
\end{figure}
Fig.~\ref{fig:trace}(a) presents per-frame median LDPC decoding times under the five conditions, where loading the LLM alone leaves the decoding time near $175$~$\mu$s.
With continuous LLM inference, the time rises to about $1.4$~ms, approximately eight times the baseline level, while that under emulated GPU loads falls in between.
It is seen that the extent to which PHY processing efficiency degrades depends on how much of the shared GPU the LLM reasoning occupies.
Note that the lower decoding time of about $0.5$ ms in frame $10$ coincides with fewer uplink transmissions, consistent with reduced LLM-PHY contention.
Fig.~\ref{fig:trace}(b) presents the cumulative distribution functions (CDFs) of all $1278$ decoding-time samples, with the $500$-$\mu$s reference line marking one radio-slot duration under the configured $30$-kHz subcarrier spacing.
At this reference line, $99.9\%$ of baseline samples but only $6.9\%$ of samples with continuous LLM inference have decoding times below $500$~$\mu$s.
These results clearly indicate that the first coupling path is far from negligible for time-sensitive PHY processing.

\section{Proposed Control-Compute Contract}
\subsection{Protocol Rules with Operator-Requirement Protection}\label{sec:contract}
\begin{figure*}[t]
\centering
\includegraphics[width=0.995\linewidth]{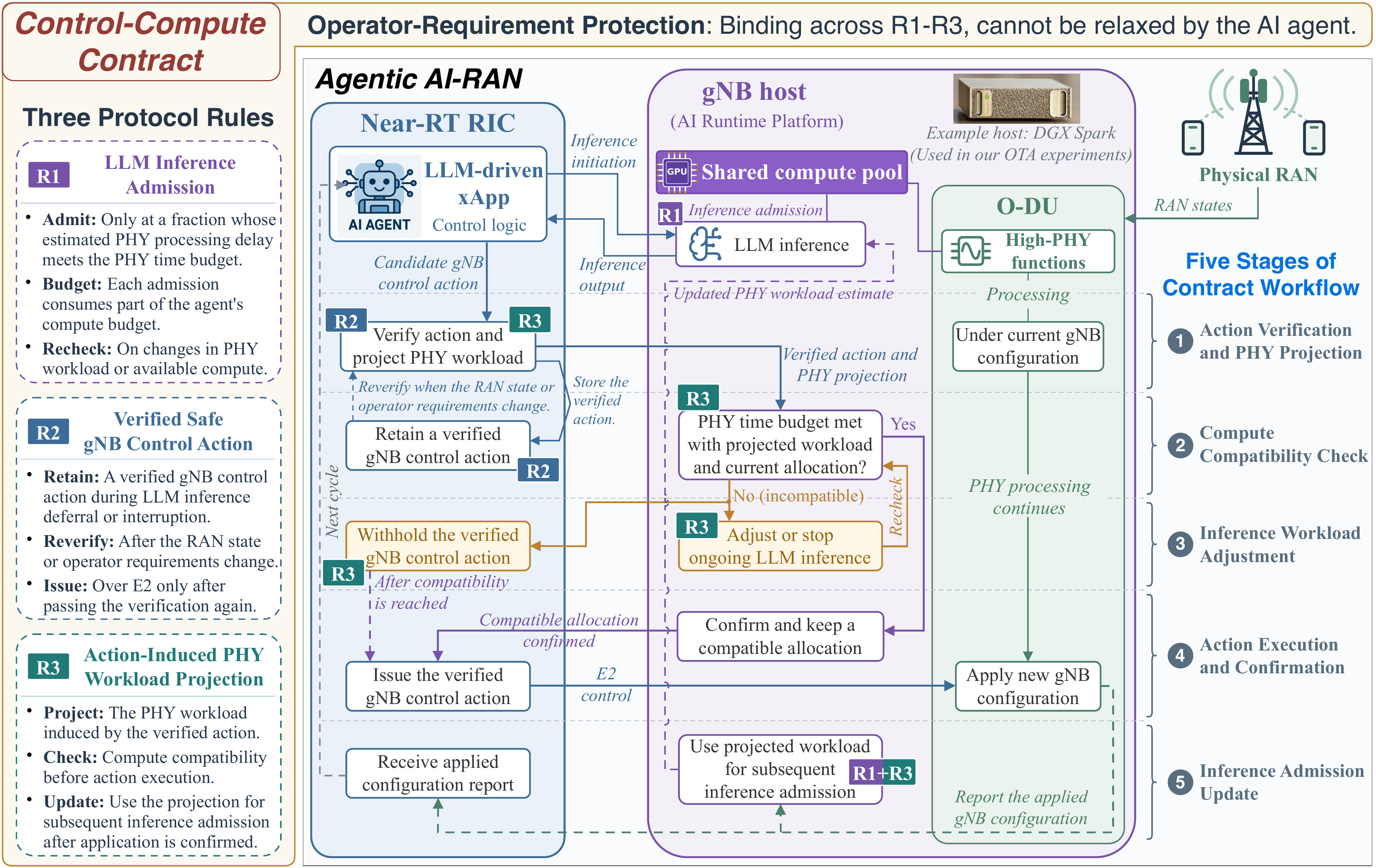}
\caption{The proposed control-compute contract, showing its three protocol rules, operator-requirement protection, and five-stage workflow within AI-RAN.}
\label{fig:epoch}
\end{figure*}

To govern the couplings identified above, we propose a control-compute contract that jointly specifies LLM inference admission, gNB control execution, and their ordering under operator requirements, as illustrated in Fig.~\ref{fig:epoch}.
Concretely, the contract consists of three protocol rules (R1 through R3) and one operator-requirement protection condition as follows.

\textbf{LLM Inference Admission (R1):}
LLM inference is admitted only at a fraction of the pool that keeps the PHY within its time budget, and each admission draws down the AI agent's remaining compute budget.
Such admission is decided at the gNB host before each inference initiation using the estimated PHY processing time under the compute allocation, and must be rechecked when the PHY workload or available compute changes.
This requires calibrating how PHY processing time varies with the fraction granted to LLM inference.

\textbf{Verified Safe gNB Control Action (R2):}
The AI agent relies on LLM inference to propose gNB control actions, but R1 may defer or interrupt that inference to protect PHY processing.
To sustain gNB control without fresh LLM outputs, R2 requires the Near-RT RIC to retain the latest verified control action, which remains issuable as long as it passes re-verification.
If the RAN state or AI-RAN operator requirements change, the retained action must pass verification again before it can be issued over E2.

\textbf{Action-Induced PHY Workload Projection (R3):}
A gNB control action verified under R2 may still change the PHY workload used for inference admission under R1.
Hence, R3 requires projecting the resulting workload before action execution and checking whether PHY processing can still meet its time budget with the compute already granted to LLM inference.
If not, the action must wait until the running inference is adjusted or stopped and a compatible compute allocation is confirmed.
Once the new configuration is confirmed as applied, its projected PHY workload must be used in subsequent inference admission under R1.

\textbf{Operator-Requirement Protection:}
Across R1 to R3, the operator requirements must remain binding and cannot be relaxed by the AI agent.
In detail, the gNB host must reject compute allocations that violate the PHY time budget or the agent's compute budget, while the Near-RT RIC must reject gNB control actions that violate the operator requirements.
Accordingly, no agent can directly obtain more compute from the shared pool by proposing gNB control actions that benefit its own inference, unless they meet the above requirements.

\textit{\textbf{Takeaway:}}
A verified gNB control action and an admitted LLM inference can still be incompatible, as the action may change the PHY workload assumed at admission.
Our contract therefore links their approval and execution order over the shared compute pool, requiring a compatible compute allocation.
On the one hand, R1 governs the first coupling path through inference admission, with R2 retaining a verified gNB control action when inference is deferred or interrupted.
On the other hand, R3 governs the second path by checking compute compatibility before action execution and updating subsequent inference admission with the projected PHY workload.

\subsection{Contract Workflow within O-RAN}
Within O-RAN, each AI agent's control logic resides in a Near-RT RIC xApp that proposes gNB control actions, as aforementioned, while its LLM inference can run on the gNB host, sharing the compute pool with accelerated PHY processing.
On the control side, the Near-RT RIC applies R2 and the R3 projection through a platform function for verified gNB control execution, which can build on an arbiter such as xTRUCE~\cite{xtruce2026}.
On the compute side, the gNB host works with the AI Runtime Platform, a software environment in the foundational layer of AI-RAN~\cite{airanarch}, to apply R1 together with the R3 compatibility check and admission update.
As depicted in Fig.~\ref{fig:epoch}, the contract workflow proceeds in five stages.

\begin{table*}[t]
\centering
\caption{Compute Allocation for LLM Inference in the Four Evaluation Cases}
\label{tab:arms}
\footnotesize
\renewcommand{\arraystretch}{1.6}
\begin{tabular}{|>{\raggedright\arraybackslash}m{1.49in}|>{\raggedright\arraybackslash}m{2in}|>{\raggedright\arraybackslash}m{1.3in}|>{\raggedright\arraybackslash}m{1.43in}|}
\hline
\textbf{Evaluation case} & \textbf{Granted $f$ in active periods} & \textbf{Granted $f$ in quiet periods} & \textbf{Quiet-to-active change during an ongoing LLM inference} \\
\hline
Ungoverned co-location (Fig.~\ref{fig:trace}) & $f=1$ & $f=1$ & Continue with $f=1$ \\
\hline
Observed-load admission~\cite{schiavo2024yyr,schiavo2024cloudric} & $f^*$ of the observed PHY workload, lagging $1$~s behind the applied gNB configuration & $f=1$ & Continue with $f=1$ \\
\hline
Worst-case reservation~\cite{sun2026sc3,yet2026slallm} & A fixed $f^*$ based on the heaviest PHY workload, i.e., $4$ code blocks per slot & The same fixed $f^*$ & Keep the fixed $f^*$ \\
\hline
\rowcolor[HTML]{F3EFF9} Proposed contract & $f^*$ based on the \textbf{projected PHY workload} once a new gNB configuration is applied & $f=1$ & \textbf{Stop the inference, then restart at $f^*$ or defer} \\
\hline
\end{tabular}
\end{table*}
\textbf{Action Verification and PHY Projection Stage:}
Whenever an LLM-driven xApp proposes a candidate gNB control action through its R1-admitted inference, the Near-RT RIC first verifies it under R2 and projects the PHY workload that would result from executing that action.
The projection result is then sent to the gNB host together with the verified action.

\textbf{Compute Compatibility Check Stage:}
Using this projection, the gNB host checks under R3 whether PHY processing can meet its required time budget with the compute already allocated to LLM inference.
If compatible, the workflow proceeds directly to action execution without adjusting inference.

\textbf{Inference Workload Adjustment Stage:}
If the check fails, the RIC withholds the new action from E2 while the host adjusts or stops any ongoing inference and rechecks the resulting compute allocation.
Meanwhile, a verified gNB control action remains available under R2, subject to re-verification when the RAN state or operator requirements change.

\textbf{Action Execution and Confirmation Stage:}
After the host confirms a compatible allocation, the Near-RT RIC issues the verified action over E2.
The host will keep that allocation in place until it confirms the O-DU has applied the new gNB configuration carried by the executed action, since an earlier change may reopen the incompatibility just resolved.

\textbf{Inference Admission Update Stage:}
Once the new gNB configuration is applied, the O-DU reports it to both the host and the Near-RT RIC, and the PHY workload projected earlier by the Near-RT RIC becomes the host's estimate for subsequent LLM inference admission under R1.
The workflow then returns to the first stage to begin the next cycle.

\section{Case Study and Preliminary Evaluation}
Consider an AI agent that serves as the uplink MCS controller of the gNB, where uplink active and quiet periods average $3$~s and $2$~s, respectively, with LDPC decoding only in active slots to trigger admission rechecks under R1.
Four MCS indices are set therein corresponding to one to four LDPC code blocks per active $0.5$-ms slot~\cite{3gpp38214}.
On the RIC side, the agent initiates an LLM inference every $30$~s, whose MCS selection is sampled with varying probabilities to sweep the mean PHY workload, over four $4000$-s runs per setting.
Here, let $f \in [0,1]$ denote the fraction of the shared compute pool granted to LLM inference.
During active periods, our contract sets $f=f^*$, the largest fraction for which the OTA-calibrated LDPC decoding time under the applied MCS remains within the illustrative $2$-ms PHY time budget with $99\%$ probability.
Accordingly, the duration of an admitted inference is approximated as the OTA-measured full-pool inference duration divided by its granted fraction.
For tractability, the operator requirements are considered met, and thus a verified MCS selection is always available under R2.
Moreover, since the selected MCS is executed only when inference completes, the compute compatibility check under R3 passes without any inference adjustment.

\begin{figure}[t]
\centering
\includegraphics[width=\linewidth]{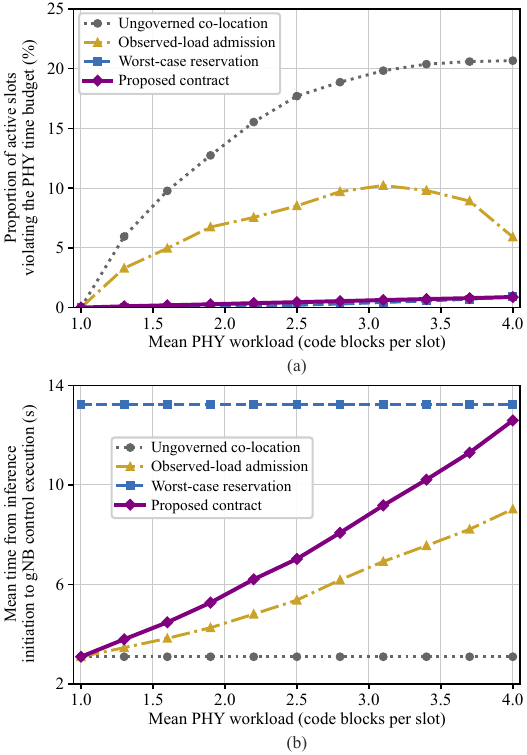}
\caption{Performance evaluation. (a) Proportion of active slots violating the PHY time budget. (b) Mean time from inference initiation to control execution.}
\label{fig:eval}
\end{figure}
Table~\ref{tab:arms} compares the proposed contract with three illustrative benchmarks, namely \textit{ungoverned co-location} (as in Fig.~\ref{fig:trace}), \textit{observed-load admission}~\cite{schiavo2024yyr,schiavo2024cloudric}, and \textit{worst-case reservation}~\cite{sun2026sc3,yet2026slallm}, which differ only in LLM inference admission and interruption.
Having these, Fig.~\ref{fig:eval}(a) shows the proportion of active slots whose decoding time exceeds the PHY time budget as the mean PHY workload varies.
Our contract stays below $0.9\%$ throughout, compared with peaks of $20.7\%$ for ungoverned co-location and $10.2\%$ for observed-load admission.
This is because the contract governs coupling by setting $f$ from the workload projected for the applied MCS and interrupting inference when its allocation becomes incompatible.
Fig.~\ref{fig:eval}(b) then presents the corresponding mean time from LLM inference initiation to gNB control execution, where the contract requires $3.1$ to $12.6$~s as the PHY workload increases, compared with $13.2$~s for worst-case reservation, i.e., up to four times faster gNB control.
This saving comes from granting inference $f^*$ of the applied MCS rather than the heaviest one, plus the full pool during quiet periods.

Table~\ref{tab:switch} further examines whether the contract's execution order can protect PHY processing in two MCS-changing cases, measured by PHY overrun slots within $2$~s after completing control action verification.
Concretely, case I raises the PHY workload from one to four code blocks per slot during an ongoing inference at $f=1$, and case II reverses this change with a new inference initiation.
Both cases assume a continuously active uplink, a $0.3$-s delay from E2 control issuance to gNB configuration application, a $0.1$-s inference stopping delay, and a $1$-s lag in observing the PHY workload.
It is observed that our contract reduces overrun slots from $2201$ to $32$ in case I by adjusting inference before the new configuration applies, whereas observed-load adjustment reacts only after the increased workload is observed.
In case II, early inference admission uses the projected workload to admit the next inference before the configuration application, yielding $597$ overruns compared with only $2$ under our contract, which defers admission until the application is confirmed.
Although these gains cost an additional $0.1$-s wait before gNB control execution in case I and a $0.3$-s deferral of the next inference in case II, our contract still efficiently governs control-compute coupling during MCS changes, which substantially reduces PHY time-budget overruns in the modeled cases.

\begin{table}[t]
\centering
\caption{PHY Time-Budget Overruns under Two Preset MCS Switches}
\label{tab:switch}
\footnotesize
\renewcommand{\arraystretch}{1.6}
\begin{tabular}{|>{\raggedright\arraybackslash}m{0.8in}|>{\raggedright\arraybackslash}m{1.21in}|>{\centering\arraybackslash}m{0.93in}|}
\hline
\textbf{Evaluation case} & \textbf{Control-compute coordination sequence} & \textbf{Expected overrun slots in $2$~s} \\
\hline
\multicolumn{3}{|l|}{\textit{I. The PHY workload rises from one to four code blocks per slot}} \\
\hline
\rowcolor[HTML]{F3EFF9} Proposed contract & Adjust inference, then apply gNB configuration & $32$ \\
\hline
Observed-load adjustment & Apply gNB configuration, then adjust inference & $2201$ \\
\hline
\multicolumn{3}{|l|}{\textit{II. The PHY workload falls from four to one code block per slot}} \\
\hline
\rowcolor[HTML]{F3EFF9} Proposed contract & Apply gNB configuration, then admit next inference & $2$ \\
\hline
Early inference admission & Admit next inference, then apply gNB configuration & $597$ \\
\hline
\end{tabular}
\end{table}

\section{Open Research Issues and Outlook}
Before agentic AI-RAN can rely on the proposed contract in real-world systems, three open issues remain as follows.

\textbf{Prolonged LLM Inference Deferral under PHY Protection:}
R1 may repeatedly defer inference to protect PHY processing, delaying the AI agent's response to RAN changes or even leaving it without an issuable action once the retained one fails re-verification.
Future work should establish conditions under which the PHY time budget remains met while the agent acts promptly.
Here, imposing a deadline on each inference response appears promising to guide compute allocation at the gNB host and support timely, verified gNB control.

\textbf{Unquantified Control Gain from Granted Compute:}
This work does not quantify how much additional LLM reasoning improves the quality of gNB control in tasks involving operator intent and changing RAN conditions.
Therefore, closed-loop evaluations with live LLM reasoning in such tasks deserve investigation to quantify the control gain per unit of granted compute and thereby to guide both the fraction granted to an agent and the spending of its compute budget.

\textbf{Calibration Drift and Stopping Delay in Practice:}
In practical O-RAN deployments, the PHY processing cost calibrated in Section~IV may drift over time, while the compute occupied by an ongoing inference is released only some time after a stop request.
This thus requires determining how much margin $f^*$ must carry against such drift and delay.
Runtime recalibration from the observed tail of PHY processing time and inference that can be paused and resumed may help maintain protection at a lower latency cost.

\section{Conclusions}
This article proposed the control-compute contract for agentic AI-RAN to address the bidirectional coupling problem arising when the AI agent simultaneously controls the gNB and utilizes the shared compute pool for its LLM inference.
Specifically, we first devised three protocol rules alongside one operator-requirement protection condition, and then presented a five-stage workflow by mapping them onto O-RAN and AI-RAN functions.
Moreover, an OTA-calibrated case study further indicated that the contract yields fewer PHY time-budget overruns than ungoverned co-location and faster gNB control than worst-case reservation, followed by several open issues and outlooks.
Looking ahead, once multiple agents share one pool and the RIC-host messages gain standardized carriers, control-compute governance can become a native part of the 6G AI-RAN control plane, treating every AI agent as both a controller and a workload.
We hope this work lays a solid foundation toward this vision of agentic AI-RAN.

\bibliographystyle{IEEEtran}
\bibliography{refs}

\end{document}